\documentclass[pra,twocolumn,aps,showpacs,floatfix,superscriptaddress]{revtex4-1}

\usepackage{graphicx}
\usepackage{rotating}
\usepackage{amsmath}
\usepackage{bbm}
\usepackage{subfigure}
\usepackage{color}

\renewcommand{\v}[1]{{\bf #1}}

\newcommand{\be}{\begin{equation}}
\newcommand{\ee}{\end{equation}}
\newcommand{\bea}{\begin{eqnarray}}
\newcommand{\eea}{\end{eqnarray}}

\graphicspath{
  {./}
  {figures/pdf/}
  {figures/eps/}
  {figures/jpg/}
}

\begin{document}

\title{Local reduced-density-matrix-functional theory: \\ Incorporating static correlation effects in Kohn-Sham equations}
\author{Nektarios~N.\ Lathiotakis}
\affiliation{Theoretical and Physical Chemistry Institute, National Hellenic 
Research Foundation, Vass.  Constantinou 48, GR-11635 Athens, Greece}
\author{Nicole\ Helbig}
\affiliation{Peter-Gr\"unberg Institut and Institute for Advanced Simulation,
Forschungszentrum J\"ulich, D-52425 J\"ulich, Germany}
\author{Angel~Rubio}
\affiliation{Nano-Bio Spectroscopy
group and ETSF Scientific Development Centre, Dpto.\ F\'isica de Materiales,
Universidad del Pa\'is Vasco, CFM CSIC-UPV/EHU-MPC and DIPC, Av.\ Tolosa 72,
E-20018 San Sebasti\'an, Spain}
\author{Nikitas~I.~Gidopoulos}
\affiliation{Department of Physics, Durham University, South Road,  
Durham DH1 3LE, United Kingdom}

\begin{abstract}  
We propose a scheme to bring reduced-density-matrix-functional theory into the 
realm of density functional theory (DFT) that preserves the accurate density functional description 
at equilibrium, while 
incorporating accurately static and left-right correlation effects in molecules and keeping the good computational performance of DFT-based schemes. The key ingredient is to relax the requirement
that the local potential is the functional derivative of the energy with respect to the density. 
Instead, we propose to restrict the search for the approximate natural orbitals
within a domain where these orbitals are eigenfunctions of a single-particle Hamiltonian with a local effective potential. 
In this way, fractional natural occupation numbers are accommodated into Kohn-Sham equations 
allowing for the description of molecular dissociation without breaking spin symmetry. 
Additionally, our scheme provides a natural way to connect an energy eigenvalue spectrum to the 
approximate natural orbitals and this spectrum is found to represent accurately 
the ionization potentials of atoms and small molecules. 
\end{abstract}

\date{\today}

\maketitle

\section{Introduction\label{sec:intro}}

Computational simulations have opened a new avenue for the
exploration and prediction of ``\`a la carte'' molecular complexes and materials,
i.e., with tailored properties and functionality, due to the development of
powerful algorithms and an increase in computational power. 
Generally, an increase in complexity or system size goes hand-in-hand with a loss of accuracy or an increase in numerical cost. 

Highly accurate wave function methods, which have recently become available also for solids \cite{BGKA2013}, are limited in the complexity of systems 
they can handle.
This hinders the application of these methods
to complex molecular structures, e.g.\ nanostructures and biomolecules. 
However, now, more than ever, there is a need for methods that are able to 
handle large-scale systems with high precision in order to assess the challenges
in material, bio-, and nanosciences.

Density functional theory (DFT)\cite{HK1964, KS1965}, on the
other hand, comes with a low computational cost, which allows its application to
rather large systems. The lack of a
systematic way to improve functionals is hampering the progress towards a more accurate
and efficient DFT first-principles scheme. 
Currently, several properties are still difficult to predict within standard DFT, with the band gap and 
molecular dissociation being prominent examples.
 The latter is described correctly with a standard local density approximation (LDA)
or generalized gradient approximation functional only if the spin symmetry is artificially broken. 
The recently introduced strictly correlated functional\cite{MG2012}  rectifies this problem,
however, at the expense of a wrong description of the equilibrium properties.
Although the exact exchange-correlation (xc) functional of DFT is universal, 
one could of course use different approximations depending on the physical situation 
or quantity of interest. To this end, large databases collecting
information from different functionals are compiled\cite{PT2013}. 
The lack of a proper description of static correlation 
is at the heart of many of the failures of present xc approximations of DFT
for describing strongly correlated systems and molecular bond breaking and formation. 
This deficiency can be
traced back to the fact that within Kohn-Sham (KS) DFT the density is reproduced via a
single Slater determinant. When the true wave function has multireference character the KS kinetic energy is a poor approximation to the true kinetic energy. The difference has to be compensated by the xc energy which for many approximations is not done very successfully, leading to a wrong total energy in the case of molecular dissociation.


In electronic structure theory, a desirable feature of independent particle models,
like the KS scheme in DFT, is the direct 
prediction of single-electron properties, like ionization potentials (IPs), from the eigenvalues 
of the single-particle Hamiltonians. 
Although the question about the physical content of the KS orbital energies raised 
a scientific debate~\cite{PY1989}, theoretical justification for this result 
was given by Baerends and co-workers~\cite{CGB2002,GBB2003} and by Bartlett and co-workers~\cite{BLS2005,VB2012}. 
Unfortunately, orbital energies from  approximate functionals tend to underestimate substantially the 
IPs of molecular systems~\cite{PN1982,PA1998}.

Reduced-density-matrix-functional theory (RDMFT) \cite{G1975} 
is an alternative to DFT to approximate the many-electron problem. It is based
on approximating the total energy of an electronic system in terms of the one-body reduced
density matrix (1RDM). A 
main difference from DFT is the introduction of fractional occupation numbers which
allows the exact treatment of the kinetic energy and potentially leads to improved 
accuracy whenever the 
ground-state many-body wave function is far from a single Slater determinant. 
Most approximations in RDMFT are explicit functionals of -- and are minimized in 
terms of -- the natural orbitals (NOs), $\phi_j(\v r)$, and their occupation
numbers, $n_j$. So far,
various different approximations for the total energy functional have become 
available\cite{M1984,GU1998,bb0,GPB2005,power_finite,pernal2010,AC3,pade,pnof1,pnof2,PNOF5,sharma08} 
which have proven to describe correctly such diverse properties as molecular  
dissociation\cite{bb0,GPB2005,power_finite,pernal2010,AC3} 
or band gaps\cite{sharma08,helbig09,dfg10,SDSG2013}. 
However, 
applications have been restricted to small molecules 
due to the computational cost to determine the orbitals. 
Although the optimization of the occupation numbers is a relatively inexpensive task,
orbital minimization is complicated: it does not reduce to an iterative eigenvalue problem 
(as in DFT) and requires numerically expensive procedures.
%
Significant effort has been devoted to devising effective
Hamiltonians\cite{pernal_epot,piris_jcc,Baldsiefen2013114} to improve the efficiency, but with
limited success so far. 

If we were able to incorporate all the advantages of RDMFT functionals into  DFT while 
keeping the cost of standard  DFT functionals we would make a big step forward in constructing 
an efficient and accurate scheme able to cope with the challenge of describing
structural and dynamical properties of many-electron systems including bond breaking and 
bond formation. In this paper, we propose such a framework that combines the best of 
both DFT and RDMFT. One can regard this approach as either an extension of DFT,
where fractional occupations for the orbitals are introduced using an
approximation for the xc energy functional borrowed from
RDMFT, or, alternatively, as a constrained RDMFT calculation. In either case, we
incorporate the proper nonidempotent nature of the density matrix in the
calculation of the kinetic energy that is fundamental to the success of
RDMFT approximations. 

The central idea in our proposed framework, which we call
{\it local RDMFT}, is to restrict the minimization with respect to the NOs
to a domain where these orbitals are eigenfunctions of a single-particle Hamiltonian with a local potential. 
The best possible Hamiltonian is the one whose eigenorbitals minimize the total energy.
The resulting equations are similar to the optimized effective potential (OEP) 
equations\cite{sharp,talman,kummel2008,BL2009}. The OEP improves the accuracy in DFT 
and significant expertise has been
developed in the past two decades\cite{kummel2008} for its efficient implementation. 
Hence, our method can be implemented directly 
in existing DFT codes with only small modifications to address fractional occupation numbers.
Fractional occupations, as in standard RDMFT, are provided by the minimization of the
energy functional under the appropriate conditions. Our approach has some similarity with 
the idea explored by Gr\"uning et al.~\cite{GGB2003}, where the common energy denominator approximation
is used together with the M\"uller functional. In that approach, occupancies are obtained in
an empirical way and not through optimization.

The local RDMFT framework provides an energy eigenvalue spectrum connected to the NOs and as we show, 
single electron properties, like the IPs of small molecules, are well reproduced by the
energies of the highest occupied molecular orbitals (HOMOs).

In Sec.~\ref{sec:theory},
we describe in detail the formalism of local RDMFT. Then in Sec.~\ref{sec:applications}, we show that
the restriction to a local effective potential has little effect on the
dissociation of dimers; hence, the accuracy of RDMFT is
retained by the proposed method. In Sec.~\ref{sec:applications} we also illustrate the quality of
the energy spectrum provided by the effective Hamiltonian by comparing the obtained IPs with
experiment. In the Appendix we show that pure density xc functionals are 
not adequate for the present scheme since they cannot lead to fractional occupations 
in a minimization procedure and we need to employ functionals of the 1RDM as we do in the present work.

\section{Local RDMFT \label{sec:theory}}

Clearly, the integration of DFT with RDMFT is desirable as it could combine the best of 
both worlds. With fractional occupation numbers, static correlation would become accessible, 
while use of a common local potential to yield the NOs would improve 
dramatically the efficiency of the method.
However, this target is not straightforward. 
A natural way to incorporate a local potential 
for the orbitals, is to consider the approximate RDMFT xc energy to be a density functional, 
so that its functional derivative would give the local potential. 
Unfortunately, such an approach leads in general to an idempotent 
density matrix (see Appendix) and we are back to square one. 

Hence, we abandon 
the requirement that the xc energy must be a functional of the density and that the potential 
must arise as a functional derivative with respect to the density.
Instead, we consider the total energy as a functional of the one-body-reduced-density matrix (1RDM), 
$\gamma ( {\bf r} , {\bf r}' ) $,
\begin{multline} \label{eq:functional}
E_v^{\rm RDM} [ \gamma ({\bf r}, {\bf r}' ) ] = T [ \gamma ({\bf r}, {\bf r}' ) ] + \int d^3 r \, \gamma ({\bf r}, {\bf r} ) v ( {\bf r}) \\ + 
\frac{1}{2} \iint d^3r d^3 r' \frac{ \gamma ({\bf r}, {\bf r} ) \gamma ({\bf r}' , {\bf r}' )}{ | {\bf r - r'} | } 
+ E_{\rm xc} ^ {\rm RDM}[ \gamma ({\bf r}, {\bf r}' )] \,,
\end{multline}
where $v( {\bf r})$ is the external local potential and $T [ \gamma ({\bf r}, {\bf r}' ) ]$ is the 
interacting kinetic energy which is a functional of the 1RDM. The 
electron-electron interaction energy can be cast into the last two terms in Eq.~(\ref{eq:functional}) 
where $E_{\rm xc} ^ {\rm RDM}[ \gamma ({\bf r}, {\bf r}' )]$ needs to be approximated.
$E_{\rm xc} ^ {\rm RDM}[ \gamma ({\bf r}, {\bf r}' )]$ can be considered a functional of the
occupation numbers $n_j$ and the natural orbitals $\phi_j$, i.e., the eigenvalues and the
eigenvectors of $ \gamma$:
\begin{equation}
\gamma ({\bf r}, {\bf r}' ) = \sum_j n_j \: \phi_j^* ({\bf r}') \: \phi_j ({\bf r})\,.
\end{equation}

The central idea in our proposed local RDMFT scheme is to restrict the search for the 
optimal $\phi_j$'s within a domain where they are also 
eigenfunctions of a single-particle Hamiltonian with a local potential, $v_{\rm rep}(\v r)$,
\begin{equation}
\label{eq:locham}
\left[ -\frac{\nabla^2}{2} + v_{\rm ext} (\v r) + v_{\rm rep}(\v r) \right] \phi_j (\v r) = 
\epsilon_j \phi_j(\v r)\,,
\end{equation}
where $v_{\rm rep}$ is the effective repulsive potential acting on any electron in the system, 
caused by the effective 
repulsion of the remaining $N-1$ electrons (atomic units are used throughout the paper).
Fractional occupations, $n_j$, as in standard RDMFT, are provided by the minimization of the
energy functional of Eq.~(\ref{eq:functional}) under the appropriate $N$-representability 
conditions. 


The natural orbitals in the exact theory, as well as the minimizing orbitals in 
RDMFT approximations, are typically satisfying a Schr\"odinger equation
with a nonlocal effective potential. The local potential constraint leads to
approximate natural orbitals (ANOs), $\phi_j$, which cannot become equal to the true natural orbitals.

By enforcing Eq.~(\ref{eq:locham}), the total energy becomes a functional of the
local effective potential and of the occupation numbers, $E=E[v_{\rm rep},\{n_j\}]$. 
In the same way as in the OEP 
method \cite{sharp,talman,kummel2008,BL2009}, the optimal local potential is obtained by 
solving the integral equation
\begin{equation} \label{oep_0}
\int d^3r' \chi (\v r , \v r') \, v_{\rm rep} (\v r') = b(\v r)\,,
\end{equation}
where $\chi (\v r, \v r')$,  a generalized density-density response function,  and $b (\v r)$ are  given by
\begin{equation}
\label{eq:A}
   \chi (\v r , \v r') = {\sum_{j \ne k}}^\prime  \phi_j^*(\v r)\,\phi_k(\v r)\,\phi_k^*(\v r')\,\phi_j(\v r')\frac{n_j-n_k}{\epsilon_j -\epsilon_k}\,, 
\end{equation}
\begin{equation}
   b(\v r) = {\sum_{j \ne k}}^\prime  \langle \phi_j | \frac{F_{\rm Hxc}^{(j)}-F_{\rm Hxc}^{(k)}}{\epsilon_j - \epsilon_k} 
| \phi_k \rangle\,  \phi_k^*(\v r)\, \phi_j(\v r)\,,
\label{eq:B}
\end{equation}
with $F_{\rm Hxc}^{(j)}$ defined by 
\begin{equation}
\label{eq:Liapis}
\frac{\delta E_{\rm Hxc}}{\delta \phi_j^*(\v r)} \doteq
\int d^3r' \, F_{\rm Hxc}^{(j)}(\v r, \v r') \, \phi_j(\v r')\,. 
\end{equation}
Here $E_{\rm Hxc}$ is the approximation for the electron-electron interaction 
energy, i.e., the last two terms in Eq.~(\ref{eq:functional}). 

In the summations of Eqs.~(\ref{eq:A}) and (\ref{eq:B}),
we have excluded terms over pairs of  orbitals differing in occupation
by less than a cutoff value $\Delta n_c$. This choice excludes pairs of weakly occupied orbitals whose
energies $\epsilon_j$ are not accurate for finite localized basis sets; 
e.g., occasionally they violate the Aufbau principle
and the negative definiteness of $\chi$. By excluding these terms, we also partly alleviate a common 
inaccuracy of most RDMFT functionals to show a spurious excess of total occupation in
weakly occupied orbitals \cite{L2013}. 
In that way, we have examined the dependence of our solution on $\Delta n_c$. Typically, for very small
values of $\Delta n_c$ we run into convergence issues and IPs vary substantially as a function of $\Delta n_c$. As
$\Delta n_c$ increases to a typical value of 0.1 convergence improves dramatically and IPs remain unchanged 
as a function of $\Delta n_c$ for a broad range of values. Even for values of $\Delta n_c$ larger than the HOMO-lowest occupied molecular orbital (LUMO) occupation difference, we do not find any noticeable change in the results.
Further increase deteriorates the accuracy since fewer terms are included in the summations.  
A choice for $\Delta n_c\sim 0.1-0.3$ usually leads to a converged solution. 
Accidental exclusions of strongly-weakly occupied pairs have a negligible effect on the results.

To ensure a physical asymptotic decay of the effective repulsive potential, we do not solve Eq. (\ref{oep_0}) directly.
Instead, we follow the methodology  in Ref.~\cite{GL2012} and express $v_{\rm rep} (\v r)$ as the electrostatic potential
of an effective repulsive density, $\rho_{\rm rep} (\v r)$: 
\begin{equation}
   v_{\rm rep}(\v r) = \int d^3r'  \frac{\rho_{\rm rep}(\v r')}{|\v r-\v r'|}\,.
\end{equation}
The requirement~\cite{GL2012} that the effective repulsive density corresponds to a (fictitious) 
system of $N-1$ electrons repelling the electron at $\v r$ yields the following two constraints:
\begin{eqnarray}
Q_{\rm rep}\doteq \int d^3r \: \rho_{\rm rep}(\v r) &=& N-1, \label{eq:asympt}\\
 \rho_{\rm rep}(\v r) & \geq & 0\,. \label{eq:pos}
\end{eqnarray}
The first condition is necessary for the asymptotic decay of the effective repulsive potential as $(N-1)/r$, which is a property of 
the exact Hxc potential\cite{PhysRevLett.83.5459}. 
The two conditions together become sufficient (although probably not necessary anymore) to guarantee the correct asymptotic 
behavior and a well-posed mathematical problem.

Minimization of the total energy leads to an integral equation for the effective repulsive density:
\begin{equation}
\int d^3 r' \, {\tilde \chi} (\v r , \v r') \, \rho_{\rm rep} (\v r') = {\tilde b}( \v r), 
\end{equation}
with
\begin{eqnarray}
\label{eq:oep2}
{\tilde \chi} ( \v r , \v r')\! \doteq\!\! \iint \!\! d^3 x \, d^3 y \,  \frac{  \chi ( \v x , \v y) }{ | \v x - \v r| | \v y - \v r'| }\,,\nonumber \\
{\tilde b} ( \v r ) \! \doteq \int \!\! d^3 x \,  \frac{ b ( \v x ) }{ | \v x - \v r|  }. 
\end{eqnarray}
The two constraints can be incorporated with a 
Lagrange multiplier (\ref{eq:asympt}) and a penalty term (\ref{eq:pos}) that introduces an energy 
cost for every point where $ \rho_{\rm rep}$ becomes negative.

We expand the ANOs in a basis set (orbital basis) and the effective repulsive density 
(rather than the potential) in another (auxiliary) basis and
we obtain, similarly to Ref.~\cite{GL2012}, a linear system of equations 
for the expansion coefficients of the repulsive density. 
The inversion of the linear equations is complicated 
by the fact that the matrix of the response function $\tilde \chi$ becomes 
singular when the auxiliary basis is large compared to the orbital 
basis\cite{SSD2006,HIGBBT2001,GL2012b,BFBG2011}.
As a result, the 
effective repulsive density becomes indeterminate in the null space of the 
(finite-orbital-basis) response function. This indeterminacy is substantially
suppressed by the two constraints, Eqs.~(\ref{eq:asympt}) and (\ref{eq:pos}), that reduce drastically the 
form of the admissible effective repulsive densities and effective repulsive potentials. 
We have found that with a singular value decomposition, we obtain systematically smooth 
and physical densities and potentials (see Ref.~\onlinecite{GL2012}).

A consequence of the local-potential approximation is that the asymptotic decay of the 
ANOs depends on the energy 
eigenvalue $\epsilon_j$ and hence, differs from the (necessarily uniform) asymptotic decay 
of exact NOs with fractional occupancy\cite{MPL1975,Davidson1976}. 
As a result, the asymptotic exponential decay of the density is related to the 
highest energy eigenvalue 
with nonzero occupation and not to the IP. 
The effect in the total energy from the different asymptotics - compared with the behavior of the exact NOs - is negligible 
since the energy contribution of the asymptotic region is insignificant.


Similarly to standard RDMFT, nominal scaling of local RDMFT is $N^4\times M$, where $M$
is the total number of different generalized Fock matrices [defined in Eq.~(\ref{eq:Liapis})] that
need to be evaluated and then transformed to obtain the matrix elements in Eq.~(\ref{eq:B}). For most approximations equal occupations
result in equal generalized Fock matrices. Hence, 
$M$ is typically equal to the number of orbitals with fractional occupations
plus an additional one common for all the fully occupied orbitals. Since $M$ can be fixed, 
the nominal scaling of local RDMFT reduces to that of Hartree-Fock (HF) or hybrid DFT calculations.
The number of different $F_{\rm Hxc}^{(j)}(\v r, \v r')$ that need to be evaluated, and the additional
cycle of convergence of the occupations apart from the ANOs add to the overall computational cost
of local-RDMFT as compared to standard DFT and Hartree-Fock calculations.  

The present scheme would benefit from the recent developments in reducing
the scaling of wave-function-based schemes~\cite{MPT2013}. Significant numerical cost saving
is also expected by improving the enforcement of the positivity condition of Eq.~(\ref{eq:pos}).
 However, the search for the optimal
repulsive density through an iterative eigenvalue equation is still much more efficient than
full minimization for the whole set of natural orbitals. This computational efficiency
represents a
big advantage of local RDMFT over standard RDMFT allowing for the first time 
relatively large systems to be in the capability of RDMFT.
The application of local RDMFT to larger systems gives very promising results\cite{our_2}. 
Finally, one advantage of this method is that it can
be easily implemented in standard electronic structure codes 
due to its similarity with the OEP method. 

Finally, with regard to 
the efficiency of our scheme, our target 
is not to replace  well-established methods in routine calculations where
these methods are known to be accurate but to complement them
and improve over their results where they deviate from experiment. 
Such cases are for instance bond breaking and highly correlated systems where static 
correlations are important. At the same time, in our scheme, the orbital energies from the 
effective Hamiltonian offer improved spectral properties.
 
\section{Applications \label{sec:applications}}

\begin{figure*}
\vspace{0.4cm}
\begin{center}
\begin{tabular}{cc}
\includegraphics [width=8cm]{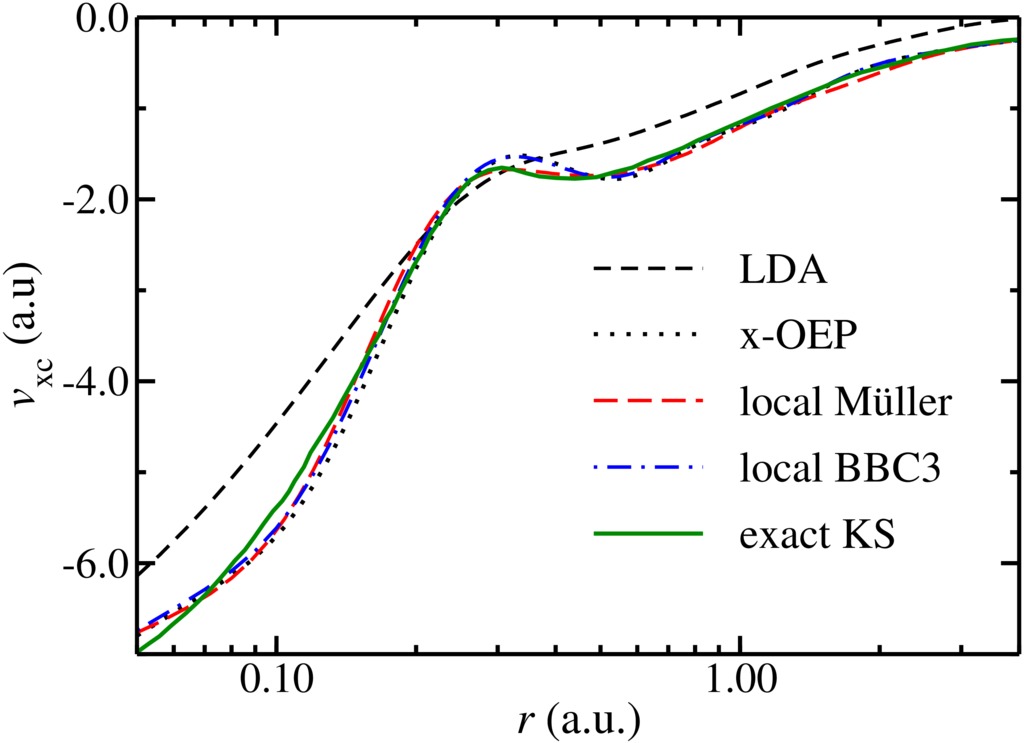} \ \ \ \ \ & \ \ \ \ \ 
\includegraphics [width=8cm]{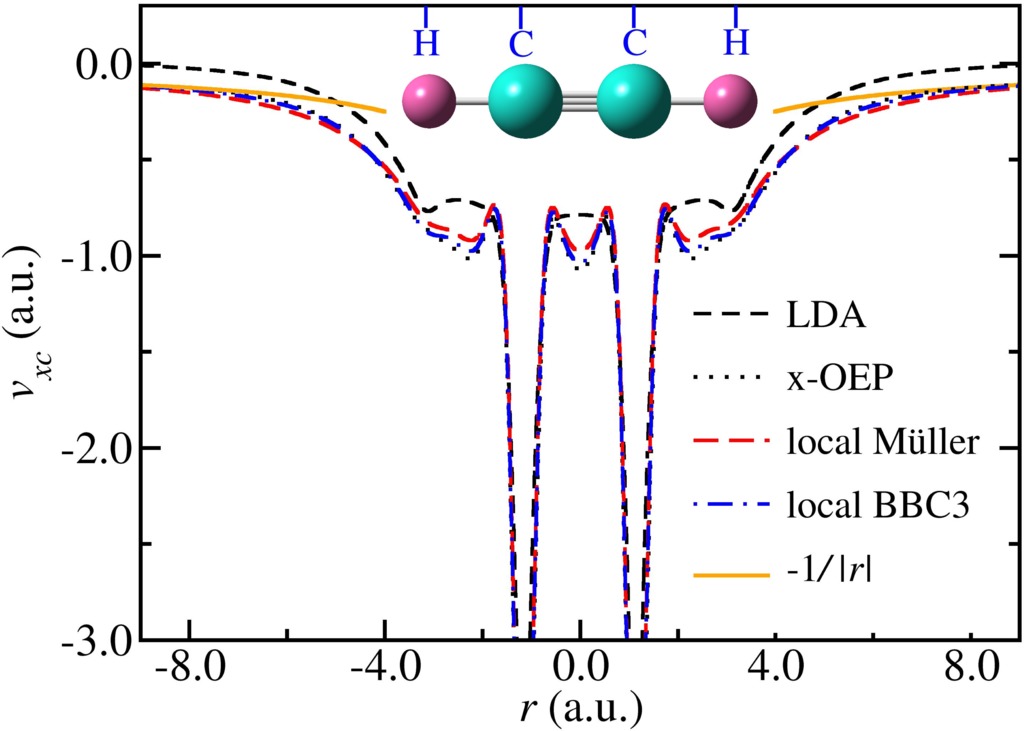} \\
\end{tabular}
\end{center}
\caption{The xc part of the optimal local potential, $v_{\rm rep}$
for the Ne atom (left) and 
acetylene molecule (right) 
using local RDMFT. For Ne,
the exact KS potential\cite{exact_KS} 
as well as the LDA and x-OEP potentials are also shown. 
\label{fig:allpots}}
\end{figure*}

In Fig.~\ref{fig:allpots}, we show the effective local potential for a Ne
atom employing two 
RDMFT approximations, the M\"uller\cite{M1984,bb0} and 
BBC3\cite{GPB2005} functionals. As we see, the optimal potentials are similar
to the exact-exchange OEP (x-OEP), especially for BBC3 while local M\"uller 
is closer to the exact KS\cite{exact_KS} potential. 
The comparison with the local potentials 
of different theories and the exact KS scheme are useful since 
 a reasonable approximate potential resembling the exact KS potential will
hopefully lead to a reasonable single-particle spectrum.

\begin{figure} 
\vspace{0.4cm}
\begin{center}
\includegraphics[width=8.0cm, clip]{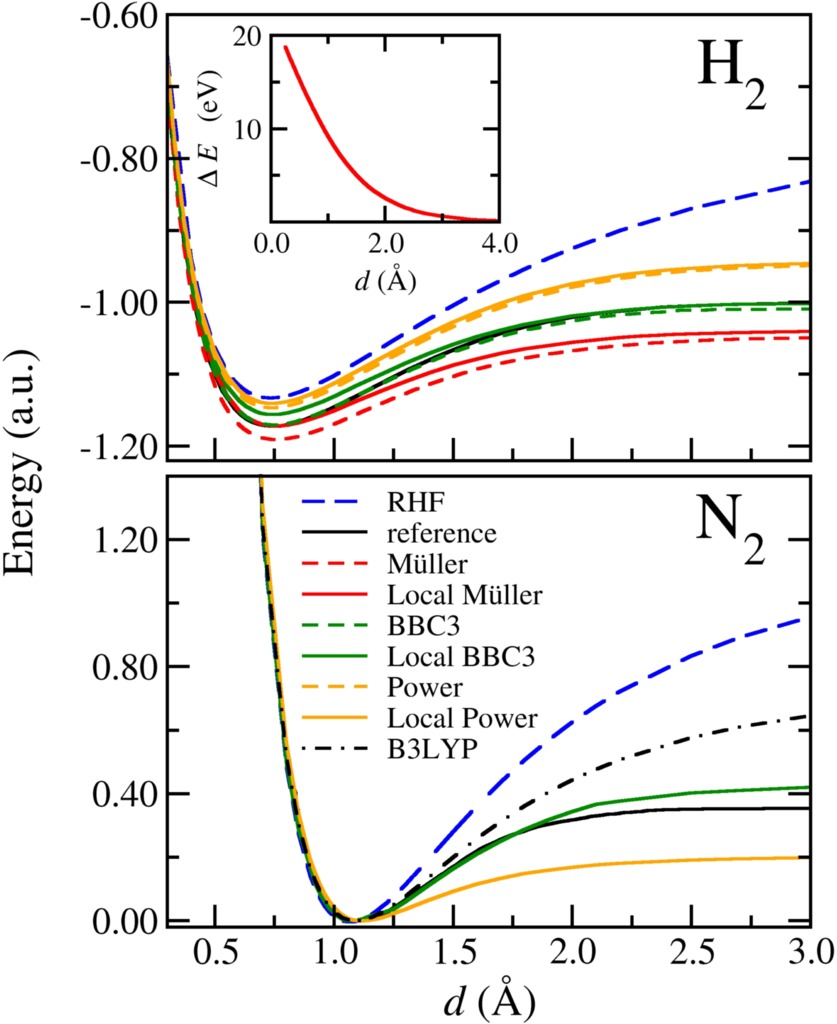} 
\end{center}
\caption{(Color online) Total energy versus internuclear separation for H$_2$ and  N$_2$ molecules. 
The HOMO-LUMO energy difference, $\Delta E$, for H$_2$, is also shown in the inset.
Energies for N$_2$ are
shifted vertically to match at the equilibrium distance. CI and experimental\cite{N2dis} results were used as 
reference for H$_2$ and N$_2$, respectively.}
\label{fig:dis}
\end{figure}

\begin{table}
\setlength{\tabcolsep}{8pt}
\begin{tabular}{lrccc}
 & Method & $R_0$ & $D_e$ (Ha) & $\omega_0$ (cm$^{-1}$) \\ \hline
H$_2$ & Local-M\"uller & 0.76 & 0.134 & 4344 \\
   & M\"uller & 0.76 & 0.143 & 4134 \\
   & Local power & 0.74 & 0.197 & 4646 \\ 
   & Power & 0.75 & 0.200 & 4277 \\
   & Local BBC3 & 0.75 & 0.155 & 4328 \\
   &  BBC3 & 0.75 & 0.156 & 4432 \\ 
   &  CI   & 0.74 & 0.172 & 4374 \\ \hline
N$_2$ & Local power & 1.10 & 0.200 & 2150 \\
      & Local BBC3  & 1.08 & 0.436 & 2500 \\ 
      & Expt. & 1.09 & 0.36 & 2360 \\ \hline
\end{tabular}
\caption{\label{tab:spec}
Equilibrium distances $R_0$,  binding energies $D_e$, and vibrational frequencies $\omega_0$,
 for H$_2$ and N$_2$ molecules
compared with CI calculations (using the same basis set), and experiment~\cite{N2dis} respectively.
For H$_2$, the full-minimization RDMFT results are also given for comparison.}
\end{table}

An important advantage of many RDMFT functionals  is the qualitatively correct dissociation
of small molecules, in contrast to available xc density functionals.
One example is the H$_2$ molecule and, as we show in 
Fig.~\ref{fig:dis}(top), this property holds for local RDMFT as well:
for all three functionals, the M\"uller, BBC3, and the power functionals,
local RDMFT reproduces the correct dissociation. The
description at the equilibrium distance also agrees very well with the
configuration interaction (CI) results~\cite{refdata,g09}
 both for the position of the minimum and
the curvature as can be seen in Table~\ref{tab:spec}. 
In the inset of Fig.~\ref{fig:dis}(top), we show
 the difference between the energy eigenvalues corresponding to the HOMO and the LUMO, 
as they are defined by the number of electrons and the assumption of a 
 single-electron picture. We find the same behavior as the one found for
 various DFT functionals in Ref.~\cite{CRHRRRS2013}, i.e., the energy difference
 is approaching zero at small distances between the two hydrogen atoms. 
In Fig.~\ref{fig:dis}(bottom), the dissociation curves for the triple bond of the N$_2$
molecule are shown for BBC3 and power functionals in comparison to HF and a DFT 
calculation~\cite{refdata,g09} using the B3LYP functional~\cite{B1993}. Unfortunately, the M\"uller 
functional fails to describe the dissociation of N$_2$.  The binding energy
 as well as the vibrational frequency obtained with BBC3 are closer 
to the experimental values than those given by the power functional. Interestingly local RDMFT results are 
qualitatively much better than  many density functional
approximations including B3LYP, as seen in Fig.~\ref{fig:dis}(bottom). It is worth mentioning 
that the correct dissociation of N$_2$ is not reproduced even at the level of accurate quantum 
chemistry methods like MP2 and single-reference coupled cluster.

\begin{table*}
\begin{center}
\begin{tabular}{ll|ccccccccc}
 System &        	& HF    	& M\"uller  & GU       & BBC3     &  AC3      & Power    & PNOF1    & ML       & Expt. IP  \\ 
\hline\hline
He &  IP 		& 24.970	& 24.69    & 24.87    & 24.57    &  24.84    & 24.84    & 24.88    & 25.15    & 24.59 \\ 
 & $E_c$     		& 0		& 0.77     & 0.75     & 0.74     &  0.75     & 0.68     & 0.75     & 0.79     &       \\ 
\hline
Be &  IP 		& 5.60  	& 9.51     & 8.46     & 8.73     &  8.46     & 8.58     & 8.44     & 8.55     & 9.32  \\ 
 &$E_c$     		& 0		& 0.87     & 0.89     & 0.90     &   0.90    & 0.91     & 0.90     & 0.94     &       \\ 
\hline
Ne &  IP1 		& 23.01   	& 22.90    & 21.32    & 20.92    &  20.88    & 21.65    & 20.91    & 21.32    & 21.60 \\ 
 &    IP2 		& 52.45   	& 46.52    & 45.02    & 44.67    &  44.59    & 45.38    & 44.62    & 45.04    & 48.47 \\ 
 &$E_c$     		& 0		& 0.75     & 0.73     & 0.70     &  0.72     & 0.65     & 0.73     & 0.84     &       \\ 
\hline
H$_2$ &  IP		& 16.17   	& 16.24    & 16.19    & 16.15    &  16.15    & 16.13    & 16.18    & 16.28    & 15.43 \\ 
 &$E_c$ 		& 0		& 0.68     & 0.67     & 0.62     &  0.65     & 0.57     & 0.65     & 0.73     &       \\ 
\hline
H$_2$O &IP1		& 13.73   	& 12.59    & 12.03    & 12.35    &  12.06    & 12.10    & 12.12    & 12.64    & 12.78 \\ 
 &      IP2		& 15.71   	& 14.21    & 14.09    & 14.42    &  14.16    & 14.06    & 14.20    & 14.75    & 14.83 \\ 
 &      IP3		& 19.15   	& 17.52    & 17.57    & 17.88    &  17.63    & 17.45    & 17.67    & 18.23    & 18.72 \\ 
 & $E_c$     		& 0		& 0.66     & 0.61     & 0.61     &  0.56     & 0.53     & 0.57     & 0.73     &       \\ 
\hline
NH$_3$  & IP1		& 11.64   	& 11.03    & 10.52    & 10.65    &  10.53    & 10.74    & 10.56    & 10.95    & 10.80 \\ 
 &        IP2		& 16.93   	& 15.22    & 15.36    & 15.48    &  15.42    & 15.39    & 15.45    & 15.88    & 16.80 \\ 
 & $E_c$     		& 0		& 0.65     & 0.59     & 0.59     &  0.54     & 0.53     & 0.55     & 0.72     &       \\ 
\hline
CH$_4$  &  IP1		& 14.82   	& 13.55    & 13.47    & 13.72    &  13.41    & 13.43    & 13.54    & 13.84    & 13.60,14.40 \\ 
        &  IP2		& 25.65   	& 21.34    & 21.16    & 21.52    &  21.20    & 21.21    & 21.32    & 21.62    & 23.00 \\ 
 & $E_c$     		& 0		& 0.62     & 0.64     & 0.55     &  0.55     & 0.46     & 0.55     & 0.71     &       \\ 
\hline
C$_2$H$_2$& IP1		& 11.07   	& 11.67    & 11.01    & 11.12    &  11.20    & 11.46    & 11.31    & 11.59    & 11.49 \\ 
 &          IP2		& 18.47   	& 16.15    & 15.98    & 16.25    &  16.29    & 16.37    & 16.39    & 16.78    & 16.70 \\ 
 &          IP3		& 20.88   	& 17.88    & 17.76    & 18.02    &  18.07    & 18.13    & 18.19    & 18.52    & 18.70 \\ 
 & $E_c$     		& 0		& 0.72     & 0.66     & 0.64     &  0.64     & 0.62     & 0.65     & 0.79     &       \\ 
\hline
C$_2$H$_4$&IP1		& 10.24   	& 10.68    & 10.43    & 10.45    &  10.61    & 10.47    & 10.59    & 10.90    & 10.68 \\ 
 &         IP2		& 13.76   	& 12.07    & 12.11    & 12.41    &  12.47    & 12.15    & 12.61    & 12.87    & 12.80 \\ 
 & $E_c$     		& 0		& 0.68     & 0.63     & 0.55     &  0.59     & 0.57     & 0.60     & 0.75     &       \\ 
\hline
CO$_2$   & IP1		& 14.74   	& 13.81    & 13.24    & 13.67    &  13.41    & 13.30    & 13.89    & 14.42    & 13.78 \\ 
         & IP2		& 19.21903	& 16.95    & 16.48    & 16.93    &  16.84    & 16.51    & 17.16    & 17.57    & 18.30 \\ 
 & $E_c$     		& 0		& 0.78     & 0.74     & 0.69     &  0.64     & 0.72     & 0.73     & 0.82     &       \\
\hline
\multicolumn{2}{l|}{$\Delta_{\rm all} $}& 7.85		& 4.06     & 5.34     & 4.01     &  4.73       & 4.56       & 4.25       &  3.11      &       \\
\multicolumn{2}{l|}{$\Delta_{\rm 1st}$}& 8.57		& 2.50     & 4.20     & 2.98     &  3.91       & 3.21       & 3.52       &  3.14      &       \\
\multicolumn{2}{l|}{ $\bar E_c$     } & -		& 0.72     & 0.69     & 0.66     &  0.65       & 0.63       & 0.67       & 0.78       &       \\
\hline
\end{tabular}
\end{center}
\caption{\label{tab:IPs} Ionization potentials (in eV) for atoms and small molecules compared to experiment, 
and the ratio, $E_c$, of correlation energy captured by the local approximation over that of
a full RDMFT minimization for several functionals. $\Delta_{\rm all} $, $\Delta_{\rm 1st}$,  are the average absolute errors for all IPs and for
the first IP, respectively, defined as $\Delta =100\times(1/N)\sum_i |(x_i - x_i^{\rm ref})/x_i^{\rm ref}|$, 
and $\bar E_c$ is the average of $E_c$ over the whole set of systems. Vertical experimental IPs in the last column are obtained from the
NIST Chemistry WebBook\cite{NIST} and references in Ref.~\cite{ZSDGL2012}.}
\end{table*}

We now focus on 
the single-electron spectrum of 
the local potential Hamiltonian.  For this 
purpose, we  obtained IPs as the corresponding eigenvalues of the local potential 
Hamiltonian for a test set of atomic and molecular systems and basis sets.
This set comprises small atomic and molecular systems 
using the cc-pVTZ and uncontracted cc-pVTZ basis sets for the orbital and
the auxiliary basis respectively, and we obtained IPs up to the third one.
Our numerical results for several RDMFT functionals are shown in Table~\ref{tab:IPs} where 
on the bottom we also show the average, 
absolute, percentage error of IPs. 
The same errors for the 
IPs  obtained as the energy eigenvalues of Hartree-Fock (Koopmans' theorem) are also included for comparison.
We find a remarkable agreement between the energy eigenvalues and the
experimental IPs for the functionals we tested.
All errors are below or around 5\%, with the Marques-Lathiotakis (ML) functional\cite{pade} 
going as low as $\sim$3\%. 
The agreement with experiment is even better for the first IPs  
with the M\"uller functional being the most accurate with an error of only $\sim$2\%. 
Overall the agreement with experimental values is very good and substantially 
better than the HF Koopmans' theorem. 

To estimate the effect of the local potential approximation we include in Table~\ref{tab:IPs}
the ratio $E_c$ of the correlation energies (defined as the energy differences from Hartree-Fock) 
with local RDMFT over those provided by the full RDMFT minimization. 
The average $E_c$ over the whole set of systems, included in the last row of Table~\ref{tab:IPs},
is in the range  0.6-0.8\% for all functionals considered.
However, as in DFT calculations, the comparison of the obtained correlation energy 
to the exact one
is not the only decisive factor to assess the accuracy of an approximation 
in reproducing many properties.
For the M\"uller functional, the local RDMFT 
recovers on average 71\% of the full minimization correlation energy. 
The M\"uller functional generally
overestimates the correlation energy substantially in the full minimization and 
the constraint of the local potential 
offers an improvement. For more accurate functionals, however, this is not always the case. 

\section{Conclusion\label{conclusion}}

In conclusion, we presented a method on how to incorporate static correlation into
KS-like equations by
employing xc functionals from RDMFT. 
We have shown that when the xc energy is a density functional then the total energy minimization 
leads to an idempotent solution (Appendix). Consequently, we relaxed the requirement that the
potential must be the functional derivative of the energy with respect to the density and 
decided to minimize the total energy with respect to the 
occupation numbers and the ANOs generated by a local effective potential.
In this way,
we manage to describe the dissociation of diatomic molecules accurately.
In addition, our approach allows us to connect a single-particle 
energy spectrum to the ANOs. This spectrum is in good agreement with experimental IPs and 
photoelectron spectra for molecules. 

The proposed method provides a powerful tool which opens a new avenue:
physically motivated approximations in density-matrix based 
schemes, able to cope with strongly correlated systems\cite{sharma08} and static correlation, can 
now be brought to the realm of DFT.  The resulting KS-like approach is able to 
simultaneously describe ground-state properties, bond breaking, and single-electron spectral properties.

The scaling of our method can be easily reduced to that of hybrid DFT methods using standard
techniques. The improved computational efficiency compared to full RDMFT minimization allows 
for the application
to large systems opening the road for an improved description of electronic 
correlations in technologically important molecular systems.

\begin{acknowledgments} 
N.N.L. acknowledges financial support from the GSRT, Greece, Polynano-Kripis project 
(Grant No. 447963), N.H. from a DFG Emmy-Noether grant, and AR from the European Community’s 
FP7 through the CRONOS project, grant agree- ment no. 280879; the European Research Council 
Advanced Grant DYNamo (ERC-2010-AdG-267374);  Grupos Consolidados UPV/EHU del Gobierno Vasco 
(Grant: IT578-13). N.I.G. thanks Professor Mel Levy for helpful comments.
\end{acknowledgments}

\appendix*
\section{A corollary from Janak's theorem}

In KS DFT the (exact or approximate) ground-state density $\rho_v$ and the ground-state energy $E_v$ of an interacting $N$-electron system in an external potential $v({\bf r})$  
are obtained by minimizing, over $N$-electron 
densities $\rho$, the expression
\begin{multline} \label{sup1}
E_{v}^{\rm KS} [ \rho ( {\bf r})] = T_s [\rho ({ \bf r} )] + \int d^3 r \, \rho ( {\bf r} ) v ( {\bf r}) + \\ 
+ \frac{1}{2} \iint d^3r d^3 r' \frac{ \rho ({\bf r}) \rho ({\bf r}') }{ | {\bf r - r'} | } + E_{\rm xc} ^{\rm KS} [ \rho ({\bf r})] \,,
\end{multline}
where $T_s [ \rho({ \bf r} )] $ is the noninteracting kinetic energy.
For our analysis, it does not matter if density-functional of the the exchange and correlation 
energy in Eq.~(\ref{sup1}) is exact or approximate, 
e.g., given by LDA, $E_{\rm xc}^{\rm LDA} [ \rho ]$. 
Strictly, only the infimum of the KS total energy functional is defined. Nevertheless, 
we assume routinely that a minimizing density exists within the space of $N$-electron, 
noninteracting $v$-representable densities, inside where we search for the minimum. 
This assumption (i.e., the existence of the minimum) amounts to restricting the domain of 
interacting $v$-representable densities under study to include only those that are also 
noninteracting $v$-representable.


On the other hand in RDMFT, the ground-state 1RDM, $\gamma ( {\bf r} , {\bf r}' ) $, of the same 
interacting system, 
and its ground-state energy $E_v$, are obtained by minimizing, 
over all $N$-representable 1RDMs $\gamma ({\bf r}, {\bf r}' )$, the functional
of Eq.~(\ref{eq:functional}).
Again, it makes no difference for our analysis if $E_{\rm xc} [ \gamma ({\bf r}, {\bf r}' )]$ is exact or an approximation.

It is desirable to bring the two theories together (KS DFT and RDMFT) 
and to obtain (approximately) the interacting, nonidempotent 1RDM $\gamma_v$ and the total energy $E_v$ by solving appropriate 
single-particle equations with a local effective potential $v( { \bf r})$. It appears this aim can be achieved easily: (a) We can 
replace the noninteracting kinetic energy in Eq.~(\ref{sup1}) by the interacting kinetic energy, written in terms of the 1RDM.
At the same time we keep the dependence of the xc energy on the density $\rho({\bf r})$, 
after correcting if necessary the xc energy functional for the kinetic part of the correlation energy.
(b) Equivalently, we can 
regard the xc energy in Eq.~(\ref{eq:functional}) as a functional of the density $ \gamma ({\bf r}, {\bf r} ) $, rather than the full 1RDM.  
Both ways amount to attempting to obtain approximately the interacting ground-state 1RDM $\gamma_v$ and the total energy $E_v$ from the minimization of a total energy expression having the following form:
\begin{multline} \label{sup3}
E_v [ \gamma ({\bf r}, {\bf r}' ) ] = T [ \gamma ({\bf r}, {\bf r}' ) ] + \int d^3 r \, \gamma ({\bf r}, {\bf r} ) v ( {\bf r}) \\ + 
\frac{1}{2} \iint d^3r d^3 r' \frac{ \gamma ({\bf r}, {\bf r} ) \gamma ({\bf r}' , {\bf r}' )}{ | {\bf r - r'} | } 
+ E_{\rm xc} [ \gamma ({\bf r}, {\bf r} )] .
\end{multline}
Compared with the KS DFT scheme, Eq.~(\ref{sup1}), where the 1RDM is constrained to be idempotent, 
the search for the 1RDM in this scheme, Eq.~(\ref{sup3}), has greater variational freedom.

Below, we show the following corollary from Janak's theorem\cite{Janak}.

{\em Excluding degeneracy of the noninteracting ground state, the minimizing $N$-representable $ \gamma_s ({\bf r}, {\bf r}' )$ of $ E_v [ \gamma ({\bf r}, {\bf r}' ) ] $ is idempotent. }

This result is unexpected, since despite the greater variational freedom compared to 
Eq.~(\ref{sup1}), the solution of Eq.~(\ref{sup3}) is still idempotent.
Also, from the corollary, it appears that employing a local potential to generate approximate NOs
for the minimization of $E^{\rm RDM}_v [ \gamma ({\bf r}, {\bf r}' ) ] $ in Eq.~(\ref{eq:functional}), would always lead to an idempotent 1RDM (unless the ground state is degenerate).
Although this is not true, it indicates the difficulty to combine a nonidempotent 1RDM with a local potential. 
The way out is to give up the requirement that the local potential must be the functional derivative of the xc energy with respect to the density.

{\em Proof}:

We must minimize $E_v [ \gamma]$ 
under the usual $N$-representability constraints for the 1RDM, i.e., that the eigenfunctions $\phi_i ({\bf r})$ of $ \gamma ({\bf r}, {\bf r}' ) $ 
are normalized and that their occupation numbers satisfy $0 \le n_i \le 1$ and $\sum_i n_i = N$. Hence, we must minimize:
\begin{equation} \label{sup4}
E_v [ \gamma ({\bf r}, {\bf r}' ) ] - \sum_i \lambda_i \int d^3 r | \phi_i ({\bf r}) |^2 - \mu \sum_i n_i
\end{equation}
where $\lambda_i, \mu$ are Lagrange multipliers to satisfy the constraints.

By varying Eq.~(\ref{sup4}) with respect to $\phi_i^*$, we obtain that the minimizing $\phi_i$ (approximate NOs) satisfy
\begin{multline} \label{sup5}
\left[ - \frac{\nabla^2}{2} + v({\bf r}) + \int d^3 r'  \frac{\gamma ({\bf r}' , {\bf r}' )} {| \bf r - r' |  } \right. \\ + 
\left. \left. \frac{\delta E_{\rm xc} [ \rho ]}{ \delta  \rho ({\bf r} ) } \right|_{\rho ({\bf r}) =  \gamma ({\bf r}, {\bf r} ) } \right] 
\phi_i ({\bf r}) = \epsilon_i \, \phi_i ({\bf r})
\end{multline}
where $\epsilon_i = \lambda_i / n_i$. \\
The optimal orbitals satisfy single-particle equations with a local potential, as intended. 
These may differ from the usual KS equations [which can be derived from Eq.~(\ref{sup1})]
since $ E_{\rm xc}^{\rm KS} [\rho]$ in Eq.~(\ref{sup1}) differs in general from $E_{\rm xc} [\rho]$ in Eq.~(\ref{sup3}).

Next, from Janak's theorem (see also the book by Dreizler and Gross,\cite{Hardy} p. 54), 
or directly, by varying Eq.~(\ref{sup4}) with respect to the occupation number $n_i$, we obtain
\begin{equation} \label{sup7}
\epsilon_i - \mu = 0\,.
\end{equation} 
Equation~(\ref{sup7}) may not hold for any $i$. It must hold when the corresponding occupation number $n_i$ is not 
equal to zero or one, i.e., for any $i$ such that $0 < n_i < 1$. 

Using {\em reductio ad absurdum}, let us assume that the $ \gamma_s ({\bf r}, {\bf r}') $ which minimizes 
$E_v [ \gamma ({\bf r}, {\bf r}')]$ is the 1RDM of an interacting wave function, exhibiting a large number of 
$n_i$ with fractional values, $0 < n_i < 1$ (an infinite number of fractional $n_i$ for a complete orbital basis). 
We are not interested in 1RDMs with just a few occupation numbers different from zero or one, which may arise when the many-body ground state is degenerate, 
since these 1RDMs do not correspond to interacting wave functions.
It follows that Eq.~(\ref{sup7}) holds for all $i$ for which $0 < n_i < 1$ and the single-particle Hamiltonian
in Eq.~(\ref{sup5}) is degenerate for all these $i$ with the same eigenvalue. 
However, it is unphysical (absurd) that the KS single-particle potential in Eq.~(\ref{sup5}), or indeed any local potential, 
can show such large degeneracy. We conclude that $ \gamma_s ({\bf r}, {\bf r}') $ cannot be the 1RDM of an interacting wave function.
QED

As with the KS total energy functional (\ref{sup1}), strictly, only the infimum of the RDMFT 
total energy functional in Eq.~(\ref{sup3}) is defined. 
For the optimization of the ANOs leading to Eq.~(\ref{sup5}), where the 
occupation numbers are held fixed, we assume once more that, for the given set of occupations 
$\{ n_i \}$, a minimizing set of orbitals exists within the space of sets of orbitals where we 
search for the minimum. 
In this space, each set of orbitals originates from a common local potential $v$ and the orbitals 
satisfy the Aufbau principle for the given set of occupations $\{ n_i \}$. 
For the special case where the fixed occupation numbers are zero or one, this assumption 
(as in the KS case) reduces to the familiar assumption that the density is non-interacting 
$v$-representable.

\bibliographystyle{apsrev}


\begin{thebibliography}{57}
\bibitem[{ref()}]{refdata}
\bibinfo{note}{Reference data used for comparison throughout this paper for
  DFT, MP2, MP4 and QCI methods were produced with the use of the Gaussian09 Rev
  A.02 program}.
\end{thebibliography}

\end{document}